\documentstyle[aps,multicol,prl,epsf]{revtex}
\begin{document}
\begin{title}
{\bf Parametric Quantum Resonances for Bose-Einstein Condensates}
\end{title}

\author{P.~G. Kevrekidis$^{1}$, A.~R. Bishop$^2$ and K.{\O}. Rasmussen$^2$}
\address{$^1$Department of Physics and Astronomy, Rutgers University
136 Frelinghuysen Rd.,  Piscataway, NJ 08854-8019,\\
$^2$ Theoretical Division, Los Alamos National Laboratory,
Los Alamos, NM 87545 \\}
\date{\today}
\maketitle

\begin{abstract}
We generalize recent work on parametric
resonances for nonlinear Schr{\"o}dinger (NLS) type equations
to the case of three dimensional Bose-Einstein condensates
at zero temperatures. We show the possibility of such
resonances in the three-dimensional case, using a moment method and
numerical simulations.
\end{abstract}

\begin{multicols}{2}
\section{INTRODUCTION}

There has been a blossoming of
literature on the features of systems exhibiting Bose-Einstein 
condensation (BEC), triggered by its 
recent experimental realization\cite{JILAMITRICEMIT2}.
Intially experiments with $10^3$ to $10^6$ atoms of
rubidium or sodium (later experiments have used lithium and eventually (spin-polarized)
hydrogen), in harmonic or cigar
shaped traps have demonstrated condensation to a ``pseudo-macroscopic'' 
level of occupancy of the ground state for $nK$ temperatures. 
Time of flight measurements, velocity distributions as well as spatial
profiles have convincingly supported the physical picture of an abrupt
transition in the behavior of the Bose gas, which has been interpreted
as the signature of BEC.

Following these experiments, many theoretical studies were launched
to characterize different aspects of Bose 
condensates such as hydrodynamic modes\cite{STR1}, 
collective excitations\cite{MIT3},
the behavior of ideal quantum fluids\cite{KET,MUL}, the fraction of
noncondensate vs. condensate atoms\cite{HUT,STR}, or the generation
and stability of vortices\cite{RIP}. In turn, experimental studies 
have progressed to address some of the theoretical 
predictions\cite{JILA2MIT4} and open up new questions.

Here, we concern ourselves with one aspect of these quantum fluids, namely parametric
driving.
For the purpose of this report, we will restrict
ourselves to the framework of the mean-field or Hartree-Fock 
approximation. This approximation is rigorously justifiable only
at $T=0$ but it is expected\cite{HUT} that the contribution of the non-condensate to the density
is quite small. It is well-known that at
this mean-field level the condensate wavefunction is governed
by the Gross-Pitaevskii (GP)\cite{GRO1PITGRO2}
equation.  An issue addressed after
the original experiments achieving the condensation was
the study of collective excitations\cite{MIT3,JILA6JILA7}. 
In these papers these excitations were induced by a harmonic trap weakly modulated in time
with appropriate types of symmetry.
More recently, it was demonstrated that extended parametric resonances can occur\cite{GR} 
in a two-dimensional (2d) NLS equation
with a harmonic trap.
This result may or may not (for reasons to be explained below) be
relevant for two dimensional studies of Bose gases. However, this 
naturally raises the question of whether a similar
result can be deduced for the $3d$ case which is certainly of direct 
relevance to experimental studies.

The main question we will address is whether
weak harmonic modulation of trapped $3d$ condensates can cause an 
anomalously large response in their wavefunction. Our answer, which will
be in the affirmative, will be motivated by mathematical analysis
using a moment method and verified by numerical simulation. 
We will briefly discuss the implications of these results and
the suggestion of relevant experiments. 

\section{MOMENT METHODS}

 Considering a spherical trap, the dimensionless GP equation for the dynamics of the BEC condensate 
is\cite{BP}
\begin{eqnarray}
i u_t=-\frac{1}{\zeta^4} \nabla^2 u+ \left (\lambda(t)r^2+\nu|u|^2\right) u.
\label{eq1}
\end{eqnarray}
Here, the subscript $t$ denotes time derivative and $\zeta=(8 \pi N|a|/a_{\perp})^{1/5}$
is a dimensionless parameter arising from the number $N$ of particles, the s-wave scattering length $a$
and from $a_{\perp}$ characterizing the strength of the 
trap (see, Ref. \onlinecite{BP}). In Eq.(\ref{eq1}) $\lambda(t)$ is a dimensionless 
function allowing for time dependence of the trap and $\nu=\mbox{sign}(a)$, generalizes the 
equation to describe attractive ($a<0)$  as well as repulsive ($a>0$)
interactions. Since, we restrict ourselves to 
spherical symmetry we only include the radial contribution in the Laplace operator.  Although we 
are mainly interested in the full three-dimensional ($3d$) case we will in general consider 
the $d$-dimensional version of Eq.(\ref{eq1}) so that
\begin{equation}
\nabla^2=\frac{1}{r^{d-1}}\frac{\partial}{\partial r}\left (r^{d-1}\frac{\partial}{\partial r}\right ).
\label{x1}
\end{equation}
Similarly to Ref. \onlinecite{GR}, we define the following quantities
\begin{eqnarray}
I_{2,a}^{(d)}&=&\int_0^\infty r^a |u|^2r^{d-1}dr,
\label{eq3}
\\
I_{3,a}^{(d)}&=&i \int_0^\infty r^a (u {u_r}^{\star}-c.c.) r^{d-1}dr,
\label{eq4}
\\
I_{4,a}^{(d)}&=&\int_0^\infty  r^a \left |\frac{\partial u}{\partial r}\right |^2 r^{d-1}dr,
\label{eq5}
\\
I_{5,a}^{(d)}&=&\int_0^\infty r^a |u|^4 r^{d-1}dr,
\label{eq6}
\end{eqnarray}
where $(d)$ indexes the dimension.
This type of nonlinear Schr{\"o}dinger (NLS) equation (\ref{eq1}) has two conserved quantities: 
as a result of the phase invariance the {\em norm} corresponding to $I_{2,0}^{(d)}$ is conserved in any dimension 
$d$. Also, since Eq.(\ref{eq1}) is a Hamiltonian system arising from 
\begin{eqnarray}
H=\int_0^\infty \left [ \zeta^{-4} | \nabla u|^2+\frac{\nu}{2}|u|^4+\lambda(t)r^2|u|^2 \right ] r^{d-1}dr
\label{ham}
\end{eqnarray} 
this quantity is conserved. It is useful to note that 
the Hamiltonian (or more appropriately {\em energy functional}), $H$,
can be expressed in terms of the moments Eqs.(\ref{eq3})-(\ref{eq6})
\begin{eqnarray}
H=\zeta^{-4}I_{4,0}^{(d)}+\frac{\nu}{2}I_{5,0}^{(d)}+\lambda(t)I_{2,2}^{(d)}.
\label{ham1}
\end{eqnarray} 

The relevance of the moments Eqs.(\ref{eq3})-(\ref{eq6}), is based on their time
evolution and in the following we will derive the relations governing this 
dynamics. The physical rationale behind such an approach lies in
the fact that the resulting equations can yield predictive diagnostics for the
dynamics of the BEC. In particular, $I_{2,2}^{(d)}$
will essentially yield the width of the spatial profile of
the wavefunction. If a condensation phenomenon is to take place
even when starting from a spatially uniform distribution (at temperatures
$T<T_c$), the width must evolve towards a  constant non-zero value.

Using Eq.(\ref{eq1}) and its complex conjugate one can derive 
\begin{eqnarray}
\label{et}
\dot I^{(d)}_{2,a}&=&\zeta^{-4}aI^{(d)}_{3,a-1},\\
\dot I^{(d)}_{3,a}&=&-4\lambda(t)I^{(d)}_{2,a+1}+4a\zeta^{-4}I^{(d)}_{4,a-1}\nonumber \\
\label{to}
&-&(a+d-1)(a-1)(a-3+d)\zeta^{-4}I^{(d)}_{2,a-3}\\
&+&\nu(a+d-1)I^{(d)}_{5,a-1}.\nonumber
\end{eqnarray}
In deriving these we assume $u$ to vanish as $r\rightarrow \infty$.

Unfortunately, it is in general impossible to close this hierarchy of equations because 
the time derivative couples to the next order e.g. $\dot I^{(d)}_{2,a}$ couples 
to $I^{(d)}_{3,a-1}$ and $\dot I^{(d)}_{3,a}$ couples to 
$I^{(d)}_{5,a-1}$ and $I^{(d)}_{4,a-1}$, and so on. 
However, combining  Eqs. (\ref{et}) and (\ref{to})
provides some insight
\begin{eqnarray}
\ddot I^{(d)}_{2,a}&=&\zeta^{-4}a\left[-4\lambda(t)I^{(d)}_{2,a}+4(a-1)\zeta^{-4}I^{(d)}_{4,a-2}\right . \nonumber \\
&-&(a+d-2)(a-2)(a-4+d)\zeta^{-4}I^{(d)}_{2,a-4}\\
\label{reduced}
&+&\nu(a+d-1)I^{(d)}_{5,a-2}\left. \right ].\nonumber
\end{eqnarray}
First, 
this confirms that the norm $I^{(d)}_{2,0}$ is conserved in any dimension.
Secondly, this relation clearly suggests $a=2$ as good choice since the 
term involving $I^{(d)}_{2,a-4}$ then vanishes irrespective of  dimension. 
Also,
this choice allows the use of Eq.(\ref{ham1}) to reduce the expression (\ref{reduced}) to
\begin{equation}
\ddot I^{(d)}_{2,2}=8\zeta^{-4}H-16\zeta^{-4}\lambda(t)I^{(d)}_{2,2}+2\nu(d-2)I_{5,0}^{(d)},
\label{virial}
\end{equation}
which corresponds to the relation commonly referred  to as the {\em virial theorem}\cite{jjr}
for the nonlinear Sch{\"o}dinger without a trap, $\lambda(t) \equiv 0$.

Clearly, the $2d$ case is special as the structure of Eq.(\ref{virial}) is
such that 
a closed time evolution can be prescribed. 
In addition, for time dependent modulation of the trap amplitude,
we find (as was observed for this problem in Ref. \onlinecite{grc} and 
again in Ref. \onlinecite{GR}) a Hill type equation which 
establishes parametric resonances for the behavior
of the width (or amplitude)  of the wavefunction.
Important as this conclusion about the $2d$ 
behavior may be for
general NLS-GP equations, it is not clear that it is relevant
to BEC. Since BEC is not possible in spatial dimensions less that three ($d<3$)\cite{MUL}
where a Kosterlitz-Thouless topological transition seems to be occurring
instead\cite{MUL,SEV}, 
the applicability of the GP equation for $d<3$ is 
controversial. Our search for condensate instabilities 
is therefore most compelling in three dimensions where no such reservations 
exist.

Although Eq.(\ref{virial}) is not closed in $3d$, it is 
easily seen that the following inequalities hold for $d \geq 2$
\begin{eqnarray}
\ddot I^{(d)}_{2,2}+16\zeta^{-4}\lambda(t)I^{(d)}_{2,2} &\leq& 8\zeta^{-4}H~~~~\mbox{for}~~~\nu<0,\\
\ddot I^{(d)}_{2,2}+16\zeta^{-4}\lambda(t)I^{(d)}_{2,2} &\geq& 8\zeta^{-4}H~~~~\mbox{for}~~~\nu>0,
\label{ineq}
\end{eqnarray}
where the equality applies to the two-dimensional case only (and in fact for the noninteracting 
$3d$ case $\nu=0$). The value of these 
inequalities lies in the predictions about the 3d case\cite{newberge}. Since we can resolve,
or at least very well characterize, the 2d behavior,
 we are now able to extend 
this to quantitative predictions about of the 3d behavior. For the attractive case $\nu<0$
the possibility of collapse occurs as $I^{(d)}_{2,2}$ can become zero in finite time.  In $2d$ and due to 
Eq.(\ref{ineq}) also in $3d$, a sufficient 
condition for collapse is $H<0$, although, depending on the 
initial configurations, collapse can be achieved even for $H>0$. 
A more complete discussion is given 
in Ref. \onlinecite{berge}. For the more realistic case
(in BEC contexts) of repulsive interaction 
$\nu>0$ we see for example that the parametric resonances that were demonstrated in the 
two-dimensional case\cite{GR} will also be present in the three-dimensional case. So, for 
instance, for $d=3$ and 
$\lambda(t)=\lambda_0^2(1+\epsilon\cos(\omega t))$ will exhibit resonance 
around $\lambda_0 =n\omega/2,~n=1,2,3..$
where the extent of the first resonance is determined by the 
inequality $|1-\omega^2/(4\lambda_0^2)| <\epsilon/2$\cite{ARN}. However, 
additional resonances may be possible
 in $3d$ due to the nonlinear driving resulting from the repulsive interaction.
Some understanding of the influence of the last term in Eq.(\ref{virial}) on the dynamics can 
be gained by assuming that the wavefunction can be approximated as 
\begin{figure}
\vspace*{290pt}

\hspace{-50pt}
\includegraphics{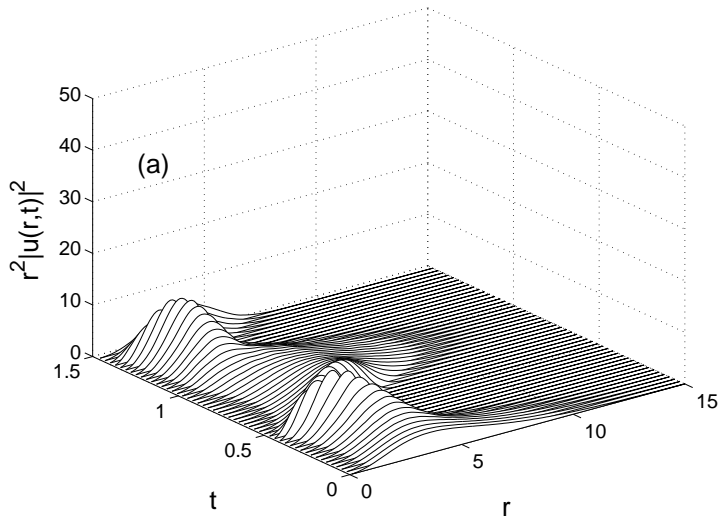}

\vspace*{200pt}

\hspace{-50pt}
\includegraphics{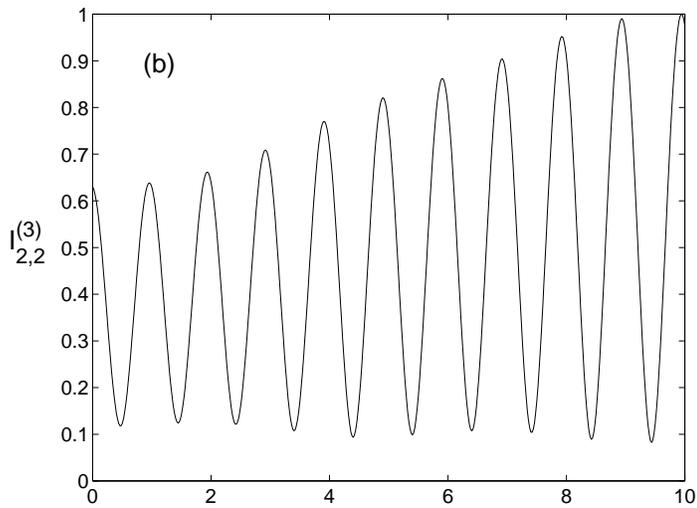}

\vspace*{-100pt}

\caption{(a) evolution (actually $r^2|u(r,t)|^2|$ is plotted for clarity)
of the condensate wavefunction both in space ($r$) and in time
($t$) for part of the domain (close to $r=0$)
and (b) the evolution if of $I^{(3)}_{2,2}$. Parameters are
$\lambda_0=1$, $\omega=1$, (i.e. second resonance) and $\epsilon=0.05$.}
\end{figure}
\begin{eqnarray}
u=\sqrt{\frac{I_{2,0}^{(3)}}{\sigma_1}}B^{-3/2}\psi(r/B),
\label{shape}
\end{eqnarray}
where $\sigma_1=\int_0^{\infty}r^2\psi(r)dr$ is a shape-dependent constant. 
Thus, assuming adiabatically the wavefunction does not alter its shape $\psi$ significantly 
as a result of the dynamics, $u$ as defined in Eq. (\ref{shape}) 
automatically 
satisfies the norm conservation. Utilizing this in Eq.(\ref{virial})
yields
\begin{eqnarray}
\frac{d^2 B^2}{dt^2}-\lambda(t)B^2=Q_1+\nu Q_2B^{-3},
\label{BS}
\end{eqnarray}
where $Q_1$ and $Q_2$ are constants determined by the 
shape $\psi$ and the initial value of $B$. Clearly, the last term in Eq.(\ref{BS}) will
only influence the dynamics when $B$ becomes small. In the repulsive $\nu>0$ case however
$B$ will generally not become small since there is no collapse. 
This simple analysis 
suggests that the parametric forcing of the experimentally 
realizable $3d$ case will result in a resonance picture analogous to
that previously reported for the $2d$ problem. 
Our numerical
simulations of the full Eq.(\ref{eq1}) with $\lambda(t)$ as given above 
verifies the validity of this prediction. A typical example for 
$\lambda_0=1$, $\omega=1$, (i.e. $n=2$) and $\epsilon=0.05$ is given in Fig.1.
The response of the wavefunction (whose initial condition
had $\mbox{max}_{x} |\psi(x,t=0)|^2=1$), corresponding to the parametric
resonance, can be observed directly from the wave function Fig. 1(a) but even
more clearly in the time evolution of $I^{(3)}_{2,2}$, as shown in Fig. 1(b).

\section{CONCLUSION AND FUTURE CHALLENGES}

In this paper, we have presented and extended the formalism
of the moment method, used in Refs. \onlinecite{GR} and \onlinecite{GR2}
for the $2d$ GP equations, to the more relevant
$3d$ case. We have commented on the special
nature of the two-dimensional problem where the moment
equations form a closed set of equations. We have also
added a note of caution in considering the results of the GP analysis
for $d < 3$. It might well be that, analogous to 
mean-field analysis in statistical physics systems, 
the ``critical dimension'' for this system
is, indeed, $d_c=3$ and for lower dimensionalities
the predictions of the mean-field theory are  unreliable.
A satisfactory self-consistent first principles description
of an interacting boson gas for $d<3$ presents
a very challenging theoretical problem, 
 and it remains an unresolved 
issue whether a transition is present (and if it is, what is
its nature).

On the other hand, we have used the moment methods 
and have derived results for
GP functionals in all dimensions of physical interest.
Considering, in particular, the  $3d$ case, where
the validity of the GP approximation is clear,
we have obtained a non-closed set of equations for the 
moments of the wavefunction.
We have demonstrated that 
for a parametric time-disturbance of the trap amplitude, 
parametric resonances are possible. 
To date the experiments that
have used parametric modulation have not observed 
such phenomena. These experiments have
been performed in cigar-shaped traps (where the analysis
is considerably more complicated even in the $2d$ 
problem\cite{GR}). No fine tuning of frequencies
and amplitudes was explored since the aim of the studies was
to excite collective modes rather than to observe parametric resonances.
Hence, we propose an experiment in a spherical trap using weak harmonic modulation of the condensate. 
Given the current experimental advances 
(see e.g. Ref. \onlinecite{STR3} for a review), such an experiment
seems feasible. Such a study
would, apart from the validation of the
theoretical prediction, also explore how such resonances might destabilize the condensate.
\section*{ACKNOWLEDGMENTS}
PGK gratefully acknowledges fellowship support from the ``A.S. Onasis''
Public Benefit Foundation and assistantship support from the Computational
Chemodynamics Laboratory of Rutgers University. 
Research at Los Alamos National Laboratory is
performed under the auspices of the US DOE.

\end{multicols}

\begin{thebibliography}{10}

\bibitem{JILAMITRICEMIT2} M.~H. Anderson, J.~R. Ensher, M.~R. Matthews, C.~E. Wieman,
and E.~A. Cornell,
{\em Science}, {\bf 269},  198 (1995);
K.~B. Davis, M.-O. Mewes, M.~R. Andrews, N.~J. van Druten, D.~S. 
Durfee, D.~M. Kurn, and W. Ketterle,
{\em Phys. Rev. Lett.}, {\bf 75},  3969 (1995);
C.~C. Bradley, C.~A. Sackett, J.~J. Tollett, and R.~G. Hulet,
{\em Phys. Rev. Lett.}, {\bf 75}, 1687 (1995);
D.~G. Freid, T.~C. Killian, L. Willmann, D. Landhuis,
S.~C. Moss, D. Kleppner, and T.~J. Greytak,
{\em Phys. Rev. Lett.}, {\bf 81},  3811 (1998).

\bibitem{STR1} A Griffin, W.C. Wu, and S. Stringari,
{\em Phys. Rev. Lett.}, {\bf 78},   1838,    1997.

\bibitem{MIT3}   M.-O. Mewes, M.~R. Andrews, N.~J. Van Druten, D.~M. Kurn,
D.~S. Durfee, C.~G. Townsend, and W. Ketterle, 
{\em Phys. Rev. Lett.}, {\bf 77}, 988, (1996).

\bibitem{KET}   W. Ketterle, and N.~J. van Druten,
{\em Phys. Rev. A}, {\bf 54},656 (1996).

\bibitem{MUL}   W.~J. Mullin,
{\em J. of Low Temp. Phys.}, {\bf 106},   615, (1997);
{\em ibid.}, {\bf 110}, 167, (1998).

\bibitem{HUT}   D.~A.~W. Hutchinson, E. Zaremba, and A. Griffin,
{\em Phys. Rev. Lett.}, {\bf 78},   1842 (1997).

\bibitem{STR} S. Giorgini, L.~P. Pitaevskii, and S. Stringari,
{\em Phys. Rev. A}, {\bf 54},   R4633 (1996).

\bibitem{RIP}   J.~J. Garcia-Ripoll, and V.~M. Perez-Garcia,
cond-mat/9903353, preprint (1999).

\bibitem{JILA2MIT4} M.~R. Mathews, B.~P. Anderson, P.~C. Haljan, C.~E. Wieman,
and E.~A. Cornell,
{\em Phys. Rev. Lett.}, {\bf 83}, 2498 (1999);
C. Raman, M. K{\"o}hl, R. Onofrio, 
D.~S Durfee, C.~E. Kuklewicz, Z. Hadzibabic, and W. Ketterle,
{\em Phys. Rev. Lett.}, {\bf 83},   2502 (1999).

\bibitem{GRO1PITGRO2} E.~P. Gross,
{\em Nuovo Cimento}, {\bf 20}, 454 (1961);
L.~P. Pitaevskii, {\em Sov. Phys. JETP}, {\bf 13}, 451 (1961);
E.~P. Gross, {\em J. Math. Phys.}, {\bf 4}, 195 (1963).

\bibitem{JILA6JILA7} D.~S. Jin, J.~R. Ensher, M.~R. Matthews,
C.~E. Weiman, and E.~A. Cornell, {\em Phys. Rev. Lett.}, {\bf 77}, 420 (1996);
D.~S. Jin, M.~R. Matthews, J.~R. Ensher,
C.~E. Weiman, and E.~A. Cornell,
{\em Phys. Rev. Lett.}, {\bf 78}, 764 (1997).

\bibitem{GR}   J.~J. Garcia-Ripoll, V.M. Perez-Garcia, and P. Torres,
{\em Phys. Rev. Lett.}, {\bf 83},   1715 (1999).

\bibitem{BP} G. Baym,  and C.~J. Pethick,
{\em Phys. Rev. Lett. }, {\bf 76},   6, (1996).

\bibitem{jjr} J.~J. Rasmussen, and K. Rypdal, Physica Scripta, {\bf 33}, 481 (1986).

\bibitem{grc} Yu.~B. Gaididei, K.~{\O}. Rasmussen and P.~L. Christiansen, Phys. Rev. E,
{\bf 52} , 2951 (1995).

\bibitem{SEV} S.~I. Shevchenkov,
{\em Sov. Phys. JETP}, {\bf 73}, 1009 (1991).

\bibitem{newberge} After the initial submission of this paper we became aware 
of a preprint (http://xxx.lanl.gov/abs/cond-mat/9907408) by L. Berge, T.~J. Alexander and Yu. ~S. Kivshar, where similar 
ideas are developed. However, Berge {\em et al.} are investigating stability 
requirements of the 3d condensates and not the possibility for parametric resonance.

\bibitem{berge} L. Berge, {\em Phys. Plasma}, {\bf 4}, 1227 (1997).

\bibitem{ARN} V.~I. Arnold,
\newblock {\it Mathematical Methods of Classical Mechanics}, (Springer-Verlag,
New York, (1989)).

\bibitem{GR2}    J.~J. Garcia-Ripoll, and V.~M. Perez-Garcia,
patt-sol/9904006, preprint,  (1999).

\bibitem{STR3}   F. Dalfovo, S. Giorgini, L.~P. Pitaevskii, and S. Stringari, 
{\em Rev. Mod. Phys.}, {\bf 71},   463 (1999).




\end{thebibliography}
\end{document}